\documentclass[12pt,a4paper]{article}
\usepackage{amssymb}
\usepackage{amsmath}

\newtheorem{theorem}{Theorem}

\newtheorem{axiom}{Axiom}

\newtheorem{corollary}{Corollary}

\newtheorem{lemma}{Lemma}

\newtheorem{proposition}{Proposition}

\input{tcilatex}

\begin{document}

\title{Measurable Systems and Behavioral Sciences\thanks{%
The first author gratefully acknowledge the support of the grant
NSh-6417.2006.6, School Support. We thank J. Busemeyer, J-C Falmagne, Y.
Feinberg, D. Luce, P. Milgrom, B. Wilson, S. Zamir, H. Zwirn as well as
seminar participants at Stanford, Irvine and Indiana for stimulating
discussions and useful comments. We are also grateful for the comments and
suggestions of three anonymous referees. } }
\author{V. I. Danilov\thanks{
CEMI, Russion Academy of Sciences Moscow, danilov@cemi.rssi.ru}\ \ and A.
Lambert-Mogiliansky\thanks{%
PSE, Paris-Jourdan Sciences, Economiques (CNRS, EHESS, ENPC, ENS) Paris\
alambert@pse.ens.fr} }
\date{}
\maketitle

\begin{abstract}
Individual choices often depend on the order in which the decisions are
made. In this paper, we expose a general theory of measurable systems (an
example of which is an individual's preferences) allowing for incompatible
(non-commuting) measurements. The basic concepts are illustrated in an
example of non-classical rational choice. We conclude with a discussion of
some of the basic properties of non-classical systems in the context of
social sciences.\ In particular, we argue that the distinctive feature of
non-classical systems translates into a formulation of bounded rationality.


\end{abstract}

\section{Introduction}

In economics, an agent is defined by her preferences and beliefs, in
psychology by her values, attitudes and feelings. One also talks about
\textquotedblleft eliciting\textquotedblright\ or \textquotedblleft
revealing\textquotedblright\ preferences and attitudes. This tacitly
presumes that those properties are sufficiently well-defined (determined)
and stable. In particular, it is assumed that the mere fact of subjecting a
person to an elicitation procedure, i.e., to \textquotedblleft
measure\textquotedblright\ her taste does not affect the taste. Yet,
psychologists are well aware that simply answering a question about a
feeling may modify a person's state of mind. For instance when asking a
person \textquotedblleft Do you feel angry?\textquotedblright\ a
\textquotedblleft yes" answer may take her from a blended emotional state to
an experience of anger. But before answering the question, it may be neither
true nor false that the person was angry. It may be a \textquotedblleft
jumble of emotions\textquotedblright \cite{Wri}. Similarly, Erev, Bornstein
and Wallsten (1993) show in an experiment that simply asking people to state
the subjective probability they assign to some event affects the way they
make subsequent decisions. The so-called \textquotedblleft disjunction
effect\textquotedblright\ (Tversky and Shafir (1992)) may also be viewed in
this perspective. In a well-known experiment, the authors find that
significantly more students report they would buy a non-refundable Hawai
vacation if they knew whether they passed the exam or failed compared to
when they don't know the outcome of the examination. In the case they
passed, some buy the vacation to reward themselves. In the case they failed,
some purchase the vacation to console themselves. When they don't know, a
seemingly inconsistent behavior is observed: fewer vacations are being
purchased than in any one of the two possible events.

In the examples above, the mere fact of subjecting an agent to a procedure
that reveals her feeling, preferences or beliefs seems to affect her. In
this paper, we propose to adopt a measurement theoretical approach to
behavior: actual behavior reveals preferences (or beliefs) in the sense of
being the outcome of a measurement of those preferences. Interestingly,
Kahneman and A. Tversky explicitly discuss some behavioral anomalies in
terms of measurement theory: \textit{\textquotedblleft Analogously, - to
classical physical measurement - the classical theory of preference assumes
that each individual has a well-defined preference order and that different
methods of elicitation produce the same ordering of
options\textquotedblright . But, \textquotedblright In these situations - of
violation of procedural invariance - observed preferences are not simply
read off from some master list; they are actually constructed in the
elicitation process.\textquotedblright } (\cite{Katver00} p. 504). A. Sen 
\cite{sen97} also emphasizes that the \textquotedblleft act of
choice\textquotedblright\ has implications for preferences. In this work we
adopt the view that performing a measurement on a system generally changes
its state. In particular, an experiment or a decision situation that reveals
a person's preferences affects that person's preferences.

Is it possible to build a predictive model of a system whose state changes
as we perform measurements on it? We assert that it is if the interaction
between systems and measurement instruments satisfies some natural
conditions. We formulate them as axioms and show that the state space is
endowed with the structure of an atomistic orthomodular orthospace and the
states are realized as probability measures on the state space.

Of course, our formalization does not build on an empty spot. The question
of modeling a system that changes when being measured is at the heart of
Quantum Mechanics (QM). Birkhoff and von Neumann's seminal article from 1936
initiated a rich literature on the mathematical foundations of QM. For an
excellent review of the field see the introductory chapter in Coecke, Moore
and Wilce (2000). Recently, the interest for QM has been rapidly expanding
to other fields. Partly, this is due to the development of quantum
computing, which inspires physicists and more recently economists to
investigate the use of quantum information in games (Eisert (1999), La Mura
(2004)). Another avenue of research has emerged in response to observations
that classical (or macro) objects (e.g. human perception or preferences) can
exhibit properties specific to QM-objects. In Lambert-Mogiliansky, Zamir and
Zwirn (2003), a Hilbert space model is proposed to describe economic agents'
preferences and decision-making. Aerts (1994), Busemeyer and Townsend (2004)
and Khrenikov et al. (2003) investigate quantum-like phenomena in
psychology. The basic idea is that the mathematical formalism of QM, often
referred to as \textquotedblleft quantum logic\textquotedblright\ rather
than its physical content, is a suitable model for describing, explaining
and predicting human behavioral phenomena in psychology and social sciences.

In this paper we expose the foundations of a general measurement theory. The
objective with the proposed formulation is to allow assessing the relevance
of this framework for social sciences\ including for the analysis of
individual choice and in particular for modelling bounded rationality.

Section 2 offers a few examples of quantum and quantum-like behavior. In
Section 3 we introduce basic notions of measurement theory, namely that of
measurement and of state. They are illustrated in models of rational choice
in Section 4. Axioms and their consequences are exposed in Sections 5 and 6.
Section 7 discusses an interpretation of the basic axioms and properties for
behavioral sciences.

\section{Examples}

\textbf{Example 1:} \textit{The spin of an electron}

An electron is endowed with several characteristics including the spin. The
spin is an intrinsic property of any particle and corresponds to a magnetic
moment which can be measured.\footnote{%
Stern and Gerlagh created an instrument such that the interaction between
the magnetic moment of the electron and that of the experimental setup
generates the splitting of a beam of electrons. A measure of the deviation
can be interpreted of the measurement of spin (along some orientation).}

It is well-known that the outcome of the measurement is always $\pm 1/2$ (in
some units) independently of the orientation of the measurement device. If
we measure a concrete electron along some axis $x$ and obtain result $+1/2$,
then a new measurement along the same axis will give the same result. Assume
we prepare a number of electrons this way. If we, for the second
measurement, modify the orientation of the axis, e.g., the measurement
device is turned by $90^{\circ }$, the result now shows equal probability
for both outcomes. As we anew perform the measurement along the $x$-axis, we
do not recover our initial result. Instead, the outcome will be $-1/2$ with
.5 probability.

We limit ourselves to noting that once the spin of the electron along some
axis is known, the results of the measurement of the spin of that electron
along some other axis has a probabilistic character. This is a central
feature. In the classic world, we are used to deal with probabilities. But
there the explanation for the random character of the outcome is easily
found. We simply do not know the exact state of the system, which we
represent by a probability mixture of other states. If we sort out this
mixture in the end we obtain a pure state and then the answer will be
determinate. In the case with the spin, it is not possible to simultaneously
eliminate randomness in the outcome of measurements relative to different
axis.\medskip

\textbf{Example 2:} \textit{A fly in a box}

Consider a box divided by two baffles into four rooms (left/front (LF),
Left/Back (LB), RF and RB. In this box, we hold a fly that flies around.
Because of the baffles, it is limited in its movements to the room where it
is.

Assume that we only have access to two types of measurements. The first
allows answering the question whether the fly is in the Left (L) or the
Right (R) half of the box. And, in the process of measurement, the baffle
between the Front (F) and the Back (B) half of the box is lifted while the
separation between Right and Left is left in place. During that process, the
fly flies back and forth from Front to Back. When the measurement operation
is over and the baffle between Front and Back put back in place, the
position of the fly is therefore quite random (LF or LB). The same applies
for the measurement of Front/Back.

Assume that we have performed the measurement L/R and obtained answer L.
Repeating that same measurement even 100 times we will always obtain the
same answer L. But if we do, in between, the F/B measurement, we have equal
(for the sake of simplicity) chances to obtain R as L. We see that the
behavior of our system reminds of that of the spin (when the Stern-Gerlach
device is rotated by an angle of $90^{\circ }$). Here the position of the
fly cannot be determined with certainty with respect to the two measurements
(LR) and (FB) simultaneously. The measurement affects the system in an
uncontrollable and unavoidable way. This simple example exhibits all basic
features of the non-classical measurement theory developed in this
paper.\medskip

\textbf{Example 3:} \textit{Attitudes and preferences}

Consider the following situation. We are dealing with a group of individuals
and we are interested in their preferences (or attitudes). We dispose of two
tests.

The first test is a questionnaire corresponding to a Prisoners' Dilemma
against an anonymous opponent. The options are cooperate (C) and defect (D).
The second test corresponds to the first mover's choice in an Ultimatum Game
(UG). The choice is between making an offer of (9,1) or of (4,6).\footnote{%
In the Ultimatum game the first mover makes an offer. The respondent either
accepts the deal and the payoffs are distributed accordingly. Or he refuses
in which case no one receives any payoff.}

The observations we are about to describe cannot be obtained in a world of
rational agents whose preferences are fully described by their monetary
payoff.\footnote{%
Game Theory uniquely predicts behavior: people defect (D) in the PD and
(with common knowledge of rationality) they offer (9,1) in UG.
Experimentalists have however taught us to distinguish between monetary
payoffs, which can be controlled and preferences, which may include features
beside monetary payoffs unknown to the designer of the experiment.} But this
is not our point. Our point is that such observations exhibit the same
patterns as the ones we described in the spin and fly example above.

Suppose that we have the following observations. The respondents who answer
C to the first questionnaire repeat (with probability close to one) their
answer when asked immediately once more. We now perform the second test (UG)
and the first test (PD) again. In that last PD test we observe that not all
respondents repeat their initial answers. A (significant) share of those who
previously chose to cooperate now chooses to defect.\footnote{%
D. Balkenberg and T. Kapplan (University of Exeter, unpublished) conducted
an experiment with those two same games but with two populations of
respondents. They investigate the frequency of the choices when the two
games are played in one order compared to when they are played in the
reverse order. The data shows an impact of the first choice on the second
which is characteristic of non-classical measurements.}

How do we understand this kind of behavior? When deciding in the PD our
respondent may feel conflicted: she wants to give trust and encourage
cooperation, but she does not like to be taken advantage of. Consider the
case when her optimistic `I' takes over: she decides to cooperate. When
asked again immediately after, her state of mind is that of the optimistic
`I' so she feels no conflict: she confirms her first choice. Now she
considers the UG. The deal (4,6) is very generous but it may be perceived as
plain stupid. The (9,1) offer is not generous but given the alternative it
should not be perceived as insulting. She feels conflicted again because her
optimistic `I' does not provide clear guidance. Assume she chooses (9,1).
Now considering the Prisoners' Dilemma again, she feels conflicted anew.
Indeed, her choice of (9,1) is not in line with the earlier optimistic mood
so she may now choose to defect.\footnote{%
We do not in any manner mean that the proposed description in terms of inner
conflict is the only possible one. A variety of psychological stories are
consistent with such phenomena of non-commutativity.}

As in the spin and the fly example, the measurement (elicitation of
preferences) affects the agent in an uncontrollable way so the observed
behavior (measurement outcomes) may exhibit instances characteristic for
quantum-like systems.

\section{Measurements and states}

In this section we introduce and discuss two basic concepts of the theory,
namely the concepts of measurement and of state.

\subsection{Measurements}

A system is anything that we can perform \textit{measurements} on. A
measurement is an interaction between a system and some measurement device,
which yields some result, the \textit{outcome of the measurement} that we
can observe and record. The set of possible outcomes of a measurement $M$ is
denoted $O(M)$. For instance in the case with the Stern-Gerlach experimental
setup, we let the electron travel through a non-homogeneous magnetic field
and observe deviation either up or down. In the example with the fly we lift
up a baffle and observe in which half of the box the fly is located. In our
third example, we let people play the Prisoner Dilemma (and the UG) and
observe their choice.

\ 

\textbf{First-kindness}

Measurements constitute a special class of interactions. We focus on
non-destructive measurements, which means that the system is not destroyed
in the process of measurement so we can perform new measurements on the
system. In particular, we can perform a measurement $M$ twice in a row. If
the outcomes of the two measurements always coincide, we say that the
measurement $M$ is a \emph{first-kind} measurement.\footnote{%
The term \textquotedblleft first-kind\textquotedblright\ measurement was
proposed by W. Pauli. J. von Neumann used the following formulation:
\textquotedblright If the physical quantity is measured twice in succession
on a system $S$ \ then we get the same value each time.\textquotedblright}

In other words the results of a first-kind measurement are repeatable
(reproducible). This is a very important point that deserves some additional
comments. One may wonder why a \textquotedblleft
measurement\textquotedblright\ would fail to satisfy the property of
first-kindness. There are several reasons for that. A first and most
important reason is that the system is evolving. For instance, the thirst of
a person running a marathon is not the same from one time to another along
the race. In this paper we focus on systems that do not have an own dynamics
(or alternatively on situations where measurements are made so close in time
that we can disregard the own dynamics). A second reason for failing
first-kindness is noise in the measurement instrument itself. We shall
assume that measurements do not bring in own uncertainty. A third reason is
that the measurement operation actually is a combination of \textit{%
incompatible}\ measurements. We return to this point soon.

In what follows, we assume that all measurements are first-kind. Indeed, if
a measurement is not first-kind it is unclear what we measure and what the
relation is between the outcome of the measurement and our system. Of
course, the question about first-kindness of any concrete measurement is an
experimental one.

\medskip

\textbf{Compatibility}

Two measurements are \textit{compatible} if they, roughly speaking, can be
performed simultaneously or more precisely, if the performance of one
measurement does not affect the result of the other. Suppose that the first
measurement gave outcome $o$; then we perform the second measurement and the
first one anew. In case we are dealing with compatible measurements we
obtain outcome $o$ with certainty.

Given two compatible measurements $M$ and $N$ we can construct a third finer
measurement. We may perform $M$ and thereafter $N$ and view this as a new
(compound) measurement $M\ast N$ with outcome set $O(M)\times O(N)$. Because
of compatibility, the measurement $M\ast N$ is a first-kind measurement.

If all measurements are compatible we can substitute them with a single
finest (complete) measurement, which is also first-kind. Performing that
measurement we learn everything about the system. Such a system is \emph{%
classical}.

The existence of incompatible measurements is a distinctive feature of
non-classical systems. It is closely related to the impact of measurements
on the state and the existence of \textquotedblleft
dispersed\textquotedblright\ states (see next subsection).

In the examples of Section 2 all measurements were incompatible.

\subsection{States}

\textbf{Measurable systems}

As we perform a measurement and observe its result we learn something about
a system. All the information that we have about a system is
\textquotedblleft encapsulated\textquotedblright\ in the \emph{state} of the
system. The state is the result of past measurements and it is the basis for
making predictions of future measurements. A theory (or a model) of a system
should describe the set of states, the results of any measurement in every
state and the change in the state induced by any measurement.\footnote{%
If the system has an own dynamic the model should be enriched with a
description of its evolution over time.}

The state of a system predicts the result of any measurement. But we do not
assume that it predicts a unique outcome. We only assume that the state
determines the probabilities for the outcomes, that is it determines a
random outcome.

In order to avoid technical subtleties associated with the notion of
probability, we shall in what follows assume that the sets $O(M)$ are
finite. In such a case, a probabilistic measure (or a random element) on $%
O(M)$ is a collection of non-negative numbers (probabilities) $\mu (o)$ for
each $o\in O(M)$ subjected to the condition $\sum_{o\in O(M)}\mu (o)=1$. The
set (a simplex indeed) of probabilistic measures on $O(M)$ is denoted $%
\Delta (O(M))$. In such a way the state $s$ defines a random outcome in $%
O(M) $, that is a point $\mu _{M}(s)\in \Delta (O(M))$ for every measurement 
$M$.

Of course, the random outcome $\mu _{M}(s)$ can be degenerated, that is $\mu
_{M}(o|s)=1$ for some outcome $o\in O(M)$. In the general case, the outcome
is random; moreover, we are interested in systems with \textquotedblleft
intrinsic uncertainty\textquotedblright . We return to this central point
later, for now we note that in the general case measurements impact on
(change) the state. Indeed, let $s$ be a state such that the outcome of a
measurement $M$ is not uniquely determined. After having performed
measurement $M$ (and obtained outcome $o$) the new state $s^{\prime }$ of
the system differs from $s$ because (according to the first-kindness of $M$)
now the result of $M$ is uniquely determined and equal to $o$.\medskip

\textbf{Definition.} A \emph{measurable system} is a system equipped with a
set $\mathcal{M}$ of first-kind measurements. A \emph{model} of a measurable
system includes the following three collections of data:

1) a set of \emph{states\ }$\mathbb{S}$ ;

2) an \emph{outcome mapping,} $\mu _{M}:\mathbb{S}\rightarrow \Delta (O(M))$
for every measurement $M\in \mathcal{M}$;

3) a \emph{transition mapping,} $\tau _{M,o}:\mathbb{S}\rightarrow \mathbb{S}
$ for every measurement $M\in \mathcal{M}$ and any of its outcome $o\in O(M)$%
.\medskip

The first mapping defines the probabilities for the possible outcomes when
performing measurement $M$ in an arbitrary state $s$. The second mapping $%
\tau _{M,o}$ points out where the state $s$ goes (transits) as we perform
measurement $M$ and obtain outcome $o\in O(M)$. We have to recognize that
the mappings $\tau_{M,o} $ are not defined for those states in which the
outcome $o $ is impossible.

It is useful at this point to introduce a few notions that we also use
later. Let $M$ be a measurement and $A\subset O(M)$. Denote 
\begin{equation*}
E_{M}(A)=\{s\in \mathbb{S},\ \mu _{M}(A|s):=\sum_{o\in A}\mu _{M}(o|s)=1\}.
\end{equation*}%
The set $E_{M}(A)$ consists of the states endowed with the following
property: the result of the measurement $M$ belongs to $A$ for sure. The set 
$E_{M}(o)$ for $o\in O(M)$ is called the \emph{eigenset} of measurement $M$
corresponding to outcome $o$.

In these terms the mapping $\tau _{M,o}$ is not defined on the subset $%
E_{M}(O(M)\setminus \{o\})$, where $o$ can not be an outcome of $M$. The
image of $\tau _{M,o}$ coincides with the eigenset $E_{M}(o)$.\medskip

\textbf{Pure states}

Although it is not necessary, we shall suppose that two states coincide if
all their predictions are the same. (Here we follow Mackey:
\textquotedblleft A state is a possible simultaneous set of statistical
distributions of the observables.\textquotedblright ). In that case, we can
consider the set $\mathbb{S}$ as some subset of the convex set $\times
_{M\in \mathcal{M}}\Delta (O(M))$.\medskip

This allows to speak about mixtures of states. A state $\sigma $ is called a
(convex or probabilistic) \emph{mixture} of states $s$ and $t$ with
(non-negative) weights $\alpha $ and $1-\alpha $, if 
\begin{equation*}
\mu _{M}(o|\sigma )=\alpha \mu _{M}(o|s)+(1-\alpha )\mu _{M}(o|t)
\end{equation*}
for any $M\in \mathcal{M}$ and any $o\in O(M)$. Mixtures of three or more
states are defined similarly. A state is said to be \emph{pure} if it is not
a non-trivial mixture of other states.

Without loss of generality one can suppose that the set of states $\mathbb{S}
$ is convex (as a subset of $\times _{M\in \mathcal{M}}\Delta (O(M))$). The
subset $\mathbb{P}$ of pure states is the set of extreme points of the
convex set $\mathbb{S}$, $\mathbb{P}=\text{ext}(\mathbb{S})$. In the sequel
we assume that $\mathbb{S}$ is the convex hull of $\mathbb{P}$, $\mathbb{S}=%
\text{co}(\mathbb{P})$. Moreover, it is quite natural to assume that the
transition mappings are linear (i.e., are compatible with the convex
structure on $S$)\textbf{.} For this reason we can work with the set of pure
states $\mathbb{P}$ instead of $\mathbb{S}$. Of course, we should keep in
the mind that the transition state $\tau _{M,o}(s)$ can be mixed.

In the classical world, pure states are \emph{dispersion-free}, that is the
outcome of any measurement performed on a system in a pure state is uniquely
determined. Randomness in the results of a measurement indicates that the
system is in a mixed state. One can sort out (or filter) this mixture by
making measurements so as to eventually obtain a pure state.

A distinctive feature of non-classical systems is the existence of dispersed
(that is non dispersion-free) pure states. This feature can be called
\textquotedblleft intrinsic uncertainty\textquotedblright . It is closely
related to two other properties of non-classical systems: the existence of
incompatible measurements and the impact of measurements on states. If a
state is dispersion-free i.e., the outcome of every possible measurement is
uniquely determined, there is no reason for the state to change. If all pure
states are dispersion-free then measurements do not impact on pure states
and therefore all measurements are compatible. On the contrary, if a state
is dispersed then by necessity it will be modified by an appropriate
measurement. On the other hand, the change in a pure state is the reason for
incompatibility of measurements. The initial outcome of the first
measurement $M$ is not repeated because the system has been modified by the
second measurement $N$.

\section{An illustration: non-classical rational choice}

Let us illustrate the above introduced notions in (thought) examples of
non-classical rational choice behavior. \medskip

We shall consider a situation where an agent is making a choice out of a set
of alternatives $X$. A primitive measurement is a choice from a subset $%
A\subset X$; the set of outcomes of the measurement is $A$ (i.e. we consider
single-valued choices). For this reason we denote such a choice-measurement $%
A$. A main idea is that a choice out of \textquotedblleft
small\textquotedblright\ subsets is well-defined and rational. By
well-defined we mean that the corresponding measurement is first-kind. By
rational we mean that consecutive choices from \textquotedblleft
small\textquotedblright\ subsets satisfy Houthakker's axiom (or the
principle of independence of irrelevant alternatives, IIA). Our motivation
is that an agent may, in his mind, structure any \textquotedblleft
small\textquotedblright\ set of alternatives, i.e., he is capable of
simultaneously comparing those alternatives. He may not be able to do that
within a \textquotedblleft big\textquotedblright\ set (which we interpret as
bounded rationality, see Section 7). This does not means that our agent
cannot make a choice from a \textquotedblleft big\textquotedblright\ set.
For example, he might use an appropriate sequence of binary comparisons and
choose the last winning alternative. However, such a compound
choice-measurement would not in general be first-kind.\medskip

We formulate Houthakker's axiom in the following way. Let $A$ and $B$ be two
\textquotedblleft small\textquotedblright\ subset, and $A\subset B$. In our
context, Houthakker's axiom consists of two parts:

1) Suppose the agent chooses from $B$ an element $a$ which also belongs to $%
A $. If the consecutive measurement is $A$ then the agent chooses $a$.

2) Suppose the agent chooses from $A$ an element $a$. If the consecutive
measurement is $B$ then the outcome of the choice is not in $A\setminus
\{a\} $.\medskip

In order to be more concrete, we shall consider as \textquotedblleft
small\textquotedblright\ subsets of the size 3 or less.\medskip

We begin with the case when all choice-measurements are compatible.
Performing binary choices we obtain some binary relation $\prec $ on $X$.
From ternary choices we see that this relation is transitive so that the
relation $\prec $ is a linear order. Therefore, it is natural to identify
the set $\mathbb{P}$ of pure states with the set of linear orders on $X$. We
obtain the well-known classical model.

We now relax the assumption of compatibility of all choice-measurements and
consider three different models.\medskip

\textbf{Model 1.} $X$ includes three alternatives $a$, $b$ and $c$. To
define a model we need to define the set of states, the outcome and the
transition mappings.

Let the set of pure states $\mathbb{P}$ consists of three states denoted $%
[a] $, $[b]$ and $[c]$. When the agent is asked to choose an item out of $X$
he chooses $a$ in state $[a]$, $b$ in $[b]$, and $c$ in $[c]$. The
choice-measurement $X$ does not change the state. When the agent is asked to
choose an item out of $\{a,b\},$ a choice-measurement that we denote by $ab$%
, he chooses $a$ in the state $[a]$, $b$ in $[b]$; in the state $[c]$ he
chooses $a$ and $b$ with equal chances (and transits into $[a]$ or $[b]$
correspondingly). Symmetrically for the choice-measurements $ac$ and $bc$.

It is clear that Houthakker's axiom is satisfied in this model. Note that
the choice measurements are incompatible. Indeed, let for instance the agent
be in the state $[c]$ and choose $c$ out of $X$. We then ask her to choose
out of $\{a,b\}$. After that choice-measurement she chooses $a$ or $b$ out
of $X$ but not $c$.\medskip

\textbf{Model 2.} The set of alternatives $X$ is the same as in the previous
model. But the set of states differs. Now we identify (pure) states with
outcomes of our four choice-measurements $ab,\ ac,\ bc,\ abc$. That is $%
\mathbb{P}=\{\underline{a}b,\ a\underline{b},\ \underline{a}c,\ a\underline{c%
},\ \underline{b}c,\ b\underline{c},\ \underline{a}bc,\ a\underline{b}c,\ ab%
\underline{c}\}$. Here $\underline{a}c$ denotes the choice of item $a$ out
of $\{a,c\}$ and so on.

To define the model we have to specify the outcomes of the measurements and
the corresponding state transition.

Let the state be $a\underline{b}c$. The outcome of measurement $abc$ is
obvious, as well as that of measurements $ab$ and $bc$ (by Houthakker's
axiom). The corresponding new states are $a\underline{b}c$, $a\underline{b}$
and $\underline{b}c$ respectively. But what about choice-measurement $ac$?
We assume that (with equal chances) the new state is $\underline{a}c$ or $a%
\underline{c}$.

Let now the state be $a\underline{b}$. By definition $b$ is the outcome of
measurement $ab$ and the state does not change. Suppose that we perform the
choice\textbf{-}measurement . We assume that the new state is $ab\underline{c%
}$ with probability 1/3 and is $a\underline{b}c$ with probability 2/3.
Houthakker's axiom says that $a$ cannot be the outcome the $abc\ $%
measurement. Suppose that the measurement $ac$ is performed. The new state
is $a\underline{c}$ with probability 2/3 and $\underline{a}c$ with
probability 1/3. Similarly, if we perform measurement $bc$ we obtain $c$
with probability 1/3 and $b$ with probability 2/3.

The outcomes and the transitions into other states are defined
symmetrically. This completes the definition of our model which obviously
satisfies the Houthakker axiom. Model 2 describes a rational (non-classical)
choice behavior as does Model 1. Yet, the pure states cannot be identified
with orderings of the alternatives as in the classical model.\medskip

Clearly, the eigensets of the measurement $abc$ are the one-element sets $\{ 
\underline{a}bc\}$, $\{a\underline{b}c\}$ and $\{ab\underline{c}\}$. The
eigensets of measurement $ab$ look more interesting. They are the
two-elements sets $\{\underline{a}b,\underline{a}bc\}$ and $\{a\underline{b}%
,a\underline{b}c\}$. The eigensets of the measurements $bc$ and $ac$ are
defined similarly. Here is the full list of the properties:

a) 3 one-element subsets: $\{\underline{a}bc\}$, $\{a\underline{b}c\}$, $\{ab%
\underline{c}\}\ $(represented by the nodes of the second line from below in
the lattice below);

b) 6 two-element subsets: $\{\underline{a}b,\ \underline{a}bc\}$, $\{a%
\underline{b},\ a\underline{b}c\}$, $\{\underline{a}c,\ \underline{a}bc\}$, $%
\{a\underline{c},\ ab\underline{c}\}$, $\{\underline{b}c,\ a\underline{b}c\}$%
, $\{b\underline{c},\ ab\underline{c}\}\ $(they\ are represented by the
nodes of the third line from below in the lattice);

c) 3 four-element subsets: $\{\underline{a}bc,\ a\underline{b}c,\ \underline{%
a}c,\ \underline{b}c\}$, $\{\underline{a}bc,\ ab\underline{c},\ \underline{a}%
b,\ b\underline{c}\}$, $\{a\underline{b}c,\ ab\underline{c},\ a\underline{c}%
,\ a\underline{b}\}$ (represented by the nodes of the fourth line of the
lattice);

d) the empty set $\emptyset $ and the whole set $\mathbb{P}$\ (respectively
the bottom and the upper node in the lattice).\medskip

Note that the intersection of properties is a property as well. The lattice
of the properties is drawn below:

\unitlength=.800mm \special{em:linewidth 0.4pt} \linethickness{0.4pt} 
\begin{picture}(81.00,64.00)(-30,5)
\put(60.00,5.00){\circle{2.00}} \put(40.00,20.00){\circle{2.00}}
\put(60.00,20.00){\circle{2.00}} \put(80.00,20.00){\circle{2.00}}
\put(57.00,30.00){\circle{2.00}} \put(63.00,30.00){\circle{2.00}}
\put(40.00,45.00){\circle{2.00}} \put(60.00,45.00){\circle{2.00}}
\put(80.00,45.00){\circle{2.00}} \put(60.00,60.00){\circle{2.00}}
\put(59.00,59.00){\vector(-4,-3){17.00}}
\put(60.00,59.00){\vector(0,-1){13.00}}
\put(61.00,59.00){\vector(4,-3){17.00}}
\put(61.00,44.00){\vector(1,-1){13.00}}
\put(59.00,44.00){\vector(-1,-1){13.00}}
\put(41.00,19.00){\vector(4,-3){18.00}}
\put(60.00,19.00){\vector(0,-1){13.00}}
\put(79.00,19.00){\vector(-4,-3){18.00}}
\put(45.00,30.00){\circle{2.00}} \put(75.00,30.00){\circle{2.00}}
\put(76.00,29.00){\vector(1,-2){4.00}}
\put(44.00,29.00){\vector(-1,-2){4.00}}
\put(50.00,30.00){\circle{2.00}} \put(70.00,30.00){\circle{2.00}}
\put(41.00,44.00){\vector(2,-3){8.67}}
\put(51.00,29.00){\vector(1,-1){8.00}}
\put(80.00,44.00){\vector(-3,-4){10.00}}
\put(69.00,29.00){\vector(-1,-1){8.00}}
\put(42.00,44.00){\vector(3,-2){20.00}}
\put(78.00,44.00){\vector(-3,-2){20.00}}
\put(56.00,29.00){\vector(-2,-1){15.00}}
\put(64.00,29.00){\vector(2,-1){15.00}}
\end{picture}

\begin{center}
Figure 1
\end{center}

\noindent We note also that the lattice is not atomistic (not all elements
can be written as the join of atoms).\medskip

\textbf{Model 3.} Here the set $X$ consists of four items: $a,b,c$ and $d$.
For any $A\subset X$ with 3 or 2 elements the corresponding
choice-measurement $A$ is first-kind. We assume as before that Houthakker's
axiom holds in any two consecutive choice-measurements. In addition, we
assume that choices out of $A$ and $B$ are compatible if $A\cup B\neq X$.
Thus, our agent is \textquotedblleft more classical\textquotedblright\ than
in Model 2.

As in the classical case we can perform binary and ternary measurements on
each triple of items taken separately and reveal a linear order on that
triple. It is therefore natural to identify the set of pure states with the
collection of those linear orders.\ There are 24 such orders: $[a>b>c]$, $%
[c>d>a]$, and so on.

Suppose that the state is $[a>b>c]$.

1) If $A\subset \{a,b,c\}$ then the outcome of choice-measurement $A$ is
determined by the order $a>b>c$; the measurement does not change the state.

2) If we perform measurement $A$ with outcome set$\ \{a,b,d\},$ the new
state will be $[a>b>d]$, $[a>d>b]$ or $[d>a>b]$ with equal chances; the
outcomes are $a$, $a$ and $d$ correspondingly. For $A=\{a,c,d\}$ or $%
\{b,c,d\}$ the new states are defined similarly.

3) If we perform measurement $A$ with outcome set $\{a,d\},$ the new state
can be one of $[a>b>d]$, $[a>c>d]$, $[a>d>b]$, $[a>d>c]$, $[d>a>b]$, and $%
[d>a>c]$ with equal chances. In the first four cases the outcome is $a$; in
the two last cases the outcome is $d$. Similarly for $A=\{b,d\}$ and $%
A=\{c,d\}$.

The eigenset of the measurement $abc$ corresponding to $a$ is $%
\{[a>b>c],[a>c>b]\}$. The eigenset of the measurement $ab$ corresponding to $%
a$ is $\{[a>b>c],\ [a>c>b],\ [a>b>d],\ [a>d>b],\ [c>a>b],\ [d>a>b]\}$%
.\medskip

Consider now a choice out of the set $\left\{ a,b,c,d\right\} .$\textbf{\ }It%
\textbf{\ }is not a first-kind choice-measurement. Here many scenarios are
possible. We shall assume the agent proceeds by making two (first-kind)
measurements. First she chooses from a pair then from the triple consisting
of the first selected item and the two remaining items. We can call this
behavior "procedural rational" because the agent proceeds as if she had
preferences over the 4 items.

Let us consider the following scenario. Assume that the agent just made a
choice in $abc$\ and that the outcome was $ab\underline{c}$\ so the state
belongs to $\{[c>b>a],[c>a>b]\}.$\ Suppose that when confronted with $abcd$\
the agent follows the following procedure she \ performs the measurement $ab$
and then the measurement $bcd$\ (this means that the first outcome is $a$%
\underline{$b$}).$\ $After the first measurement,\ the state can be anyone
of $\left[ c>b>a\right] $,$\ \left[ b>a>d\right] ,\ \left[ b>d>a\right] \ $%
and $\left[ d>b>a\right] .\ $Therefore\ the outcome of $bcd$\ can be $b.\ $%
This violates of the principle of independence of irrelevant alternative
(IIA) and demonstrates preference reversal.\footnote{%
In models 1 and 2 we could also obtain a phenomena of preference reversal
but not in two consecutive choices. Model 3 allows for that because the
choice out of four items is a compound measurement.}\textbf{\ }

The examples above demonstrate that there can be many different models of
one and same measurable system. Which of them is the correct one? It is an
empirical question. We also see from this illustration that as we consider
the possibility of incompatible choice-measurements on subsets of $X$, the
behavior of a non-classical rational man differs from that of a classical
rational man in ways that can accommodate behavioral anomalies.

\section{Basic structure on the state space}

In this section we go further with the formal investigation. We show that if
the model of a measurable system ($\mathcal{M},\mathbb{S},\mu ,\tau )$ is
endowed with some additional properties (that we formulate as axioms) then
the set of states is equipped with the structure of an orthomodular
ortho-separable orthospace.

\subsection{Properties}

We remind that, for $M\in \mathcal{M}$ and $A\subset O(M)$, we introduced
the set $E_{M}(A)$ as the set of states $s$ such that when performing
measurement $M$ on a system in state $s$, the result of the measurement
belongs to $A$ for sure. The sets of the form $E_{M}(A)$ are called \emph{%
properties} of our system, namely the property to imply $A$ when performing
measurement $M$.

Different instruments measure different properties of a system. But it may
happen that one and the same property can be measured by several
instruments. We require that in such a case the probability for the property
does not depend on the instrument. We note that Model 1 from Section 4 does
not meet that requirement. Indeed, property $\left[ a\right] $ can be
obtained by measurements $abc$ and $ab.\ $But in state $\left[ c\right] ,$
outcome $a$ never obtains with the first instrument while $a$ obtains with
probability 1/2 with the second instrument.\ Yet, this is a very reasonable
requirement and we impose the following (slightly stronger) monotonicity
condition:

\begin{axiom}
Let $M$ and $M^{\prime }$ be two measurements, let $A\subset O(M)$, and $%
A^{\prime }\subset O(M^{\prime })$. Suppose that $E_{M}(A)\subset
E_{M^{\prime }}(A^{\prime })$. Then $\mu _{M}(A|s)\leq \mu _{M^{\prime
}}(A^{\prime }|s)$ for any state $s$.
\end{axiom}

In particular, if $P=E_{M}(A)=E_{M^{\prime }}(A^{\prime })$ is a property
then the probabilities $\mu _{M}(A|s)$ and $\mu _{M^{\prime }}(A^{\prime
}|s) $ are equal and depend only on the property $P$. We denote this number
as $s(P)$ and understand it as the probability to obtain property $P$ when
performing a suitable measurement of the system in state $s$. By definition
we have 
\begin{equation*}
s(P)=1\Leftrightarrow s\in P.
\end{equation*}

The set of all properties is denoted by $\mathcal{P}$. As a subset of $2^{%
\mathbb{S}}$, $\mathcal{P}$ is a poset (a partially ordered set). It has a
minimal element $\mathbf{0}=\emptyset $ and a maximal element $\mathbf{1}=%
\mathbb{S}$. Any state $s\in \mathbb{S}$ defines the monotone function $s:%
\mathcal{P}\rightarrow \mathbb{R}_{+}$, $s(\mathbf{0})=0$, $s(\mathbf{1})=1$%
.\medskip

We now describe another basic structure on the property poset $\mathcal{P}$.
Let $P=E_{M}(A)$ be a property. The subset $P^{op}=E_{M}(O(M)\setminus A)$
is a property too. We have $s(P^{op})=1-s(P)$ for any state $s\in \mathbb{S}$%
. Thus 
\begin{equation*}
P^{op}=\{s\in \mathbb{S},\ s(P^{op})=1\}=\{s\in \mathbb{S},\ s(P)=0\}
\end{equation*}%
and the set-property $P^{op}$ depends only on $P$. We call it the \emph{%
opposite} property to $P$.

\begin{lemma}
The poset $\mathcal{P}$ equipped with the operation ' is an orthoposet. That
is the following assertions holds:

1) If \ $P\subset Q\ $then$\ Q^{op}\subset P^{op};$

2) $P\cap P^{op}=\varnothing $ for every property $P$;

3) $P^{^{\prime \prime }}=P$ for every property $P$.
\end{lemma}

\textbf{Proof.} 1) Suppose $s\in Q^{op}.\ $Then\ $s\left( Q^{op}\right) =1$
and $s\left( Q\right) =1-s\left( Q^{op}\right) =0.$ Since $P\subset Q$ by
Axiom 1 we have that $s\left( P\right) =0$ as well. Therefore $s\left(
P^{op}\right) =1-s\left( P\right) =1$ and $s\in P^{op}$. Assertions 2) and
3) are obvious. $\Box$\medskip

From Lemma 1 we obtain that $\mathbf{1}$ is the only property which contains
both $P$ and $P^{op}$. Indeed if $P\cup P^{\prime }\subset Q$ then by 1) $%
Q^{op}\subset P^{op}\cap P^{\prime \prime }=\varnothing ,\ Q^{op}=\mathbf{0}$
and $Q=\mathbf{1.\ }$In other words the supremum $P\vee P^{op}$ equals$\ 
\mathbf{1}$ in the \emph{ortho-poset} $\mathcal{P}$. Note that in general $%
P\cup P^{op}\neq \mathbb{S}$.

\subsection{Orthospaces}

The set of states $\mathbb{S}$ possesses a similar orthogonality structure.
We say that two states $s$ and $t$ are \textit{orthogonal} (and write $%
s\perp t$) if there exists a property $P$ such that $s(P)=1$ and $t(P)=0$.
Since, for the opposite property $P^{op}$, it holds $s(P^{op })=0$ and $%
t(P^{op})=1$, we have $t\perp s$, so that $\perp $ is a symmetric relation
on the set $\mathbb{S}$. Clearly, $\perp $ is an irreflexive relation.

This lead us to the following general notion.\medskip

\textbf{Definition.} A symmetric irreflexive binary relation $\bot $ on a
set $X$ is called an \emph{\ orthogonality relation}. A set $X$ equipped
with an orthogonality relation $\bot $ is called an \emph{orthospace}%
.\medskip

\textbf{Example 4.} Consider an Euclidean space $H$ equipped with a scalar
product $\left( x,y\right) .\ $We say that vectors $x$ and $y$ are
orthogonal if $(x,y)=0$. The symmetry of the orthogonality relation follows
from the symmetry of the scalar product. To obtain the irreflexivity we have
to remove the null vector. So that $X=H\backslash \{0\}$ is an orthospace. A
Hilbert space over the field of complex numbers is another example. Such a
model is standard for Quantum Mechanic.\medskip

When graphically representing an orthospace, one may connect orthogonal
elements and obtain a (non-oriented) graph. This representation is often the
most \textquotedblleft economic\textquotedblright , since few edges need to
be written. Alternatively, one may connect non-orthogonal ({or \textit{%
tolerant}\footnote{%
The term \textquotedblleft tolerant\textquotedblright\ is used in
mathematics to refer to a symmetric and reflexive relation.}) elements. The }%
tolerance{\ graph quickly becomes extremely complex. In simple cases, we
combine the two representations. Dotted lines depict orthogonality and solid
lines depict tolerantness. The graphs of Example 2 (A fly in a box) is in
figure 2. }

\begin{eqnarray*}
&&\FRAME{itbpF}{1.8308in}{1.6259in}{0in}{}{}{qlbfig3.png}{\special{language
"Scientific Word";type "GRAPHIC";display "USEDEF";valid_file "F";width
1.8308in;height 1.6259in;depth 0in;original-width 2.8643in;original-height
2.4448in;cropleft "0";croptop "1";cropright "1";cropbottom "0";filename
'/document/graphics/qlbfig3.png';file-properties "XNPEU";}%
} \\
&&\ \ \ \ \ \ \ \ \ \text{Figure\ 2}
\end{eqnarray*}

For $A\subset X$, we denote by $A^{\perp }$ the set of elements that are
orthogonal to all elements of $A$, 
\begin{equation*}
A^{\perp }=\{x\in X,\ x\perp A\}.
\end{equation*}%
For instance $\emptyset ^{\perp }=X$ and $X^{\perp }=\emptyset $. If $%
A\subset B$ then $B^{\perp }\subset A^{\perp }$.\medskip

\textbf{Definition.} The subset $A^{\perp \perp }$ is called the \emph{%
ortho-closure} of a subset $A$. A set $F$ is said to be \emph{ortho-closed}
(also called a \emph{flat }) if $F=F^{\perp \perp }$.\medskip

It is easily seen that for any $A\subset X,\;$the set$\;A^{\perp }$ is
ortho-closed. Indeed, let $F=A^{\perp }$. Then $F\subset F^{\perp \perp }$.
On the other side, $A \subset A^{\perp \perp }=F^{\perp }$; applying $\perp $
we reverse the inclusion relation $F^{\perp \perp }\subset A^{\perp }=F$. In
particular, the ortho-closure of any $A$ is ortho-closed.\medskip

Let $\mathcal{F}(X,\perp )$ denote the set of all flats of orthospace $%
(X,\bot)$ ordered by the set-theoretical inclusion, which we denote $\leq $.
It contains the largest element $X$, denoted $\mathbf{1}$, and the smallest
element $\emptyset $, denoted $\mathbf{0}$. Moreover the poset $\mathcal{F}%
\left( X,\perp \right) $ is a (complete) lattice. The intersection of two
(or more) flats is a flat implying that $A\wedge B$\ exists and equals $%
A\cap B$. The join $A\vee B$\ also exists and is given by the formula%
\begin{equation*}
A\vee B=\left( A\cup B\right) ^{\perp \perp }.
\end{equation*}

\textbf{Definition.} An \emph{ortholattice} is a lattice equipped with a
mapping $\perp :\mathcal{F} \rightarrow \mathcal{F} $ such that

i. $x=x^{\perp \perp }$;

ii. $x\leq y$ if and only if $y^{\perp }\leq x^{\perp }$;

iii. $x\vee x^{\perp }=\mathbf{1}$.\medskip

Thus the poset $\mathcal{F}(X,\perp )$ is an ortholattice.

\subsection{The intersection axiom}

We have associated to a measurable system two objects: the orthoposet of
properties $\mathcal{P}$ and the ortholattice of flats $\mathcal{F}(\mathbb{S%
},\bot )$. It is intuitively clear that these two objects are closely
related. But for now we can only assert the following inclusion 
\begin{equation*}
P^{op}\subset P^{\bot }
\end{equation*}%
for a property $P$. Indeed, by the definition of $\bot $, any element of $P $
is orthogonal to any element of the opposite property $P^{op}$.

In order to go further we impose the following \textit{intersection axiom}

\begin{axiom}
The intersection of any properties is a property.
\end{axiom}

Axiom 2 puts conditions on the set of measurements as it requires that if $P$
and $Q$\thinspace are two properties there must exist a measurement such
that $P\cap Q$ is one of its eigensets. Axiom 2 is fulfilled in Models 2 and
3 from Section 4. As we shall see this Axiom implies that properties and
flats are the same. To prove this we first get a few consequences of Axiom 2.

\begin{lemma}
$P^{op}=P^{\bot }$ for any property $P$.
\end{lemma}

\textit{Proof}. Since $P^{op}\subset \mathbf{P}^{\bot }$, we have to check
the opposite inclusion $P^{\bot }\subset P^{op}$. Let $t$ be an arbitrary
element of $P^{\bot }$, and let $s$ be an arbitrary element of $P$. Since $%
t\bot s$, then by the definition of $\bot $ there exists a property $E_{s}$
such that $t\in E_{s}$ and $s\in E_{s}^{op}$. Set $E=\cap _{s\in P}E_{s}$;
by Axiom 2 $E$ is a property. Since $E\subset E_{s}$ for any $s$, we have $%
E_{s}^{^{op} }\subset E^{op}$. Together with $s\in E_{s}^{op}$ we obtain
that $P\subset E^{op}$. Therefore $E\subset P^{op}$, and we have that $t\in
\cap _{s}E_{s}=E\subset P^{op}$. $\Box $\medskip

In particular, any property $P=(P^{op})^{op}=(P^{op})^\bot$ is orthoclosed.
Hence the inclusion $\mathcal{P}\subset \mathcal{F}$ holds. We now prove the
inverse inclusion, that is any flat is a property. We first establish this
faci for flats of the form $\{s\}^{\bot }$, where $s$ is a state. Let $P(s)$
denote the least property containing the state $s$, that is intersection of
all properties containing the state $s$.

\begin{lemma}
{$s^{\bot }=P(s)^{op}$. }
\end{lemma}

\textit{Proof.} Since $s\in P(s)$, we have that $P(s)^{op }=P(s)^{\bot }$
(by Lemma 1) is contained in $\{s\}^{\bot }$. In order to check the reverse
inclusion, we consider an arbitrary element $t$ of $\{s\}^{\bot }$. By
definition this means that $s\in E$ and $t\in E^{op }$ for some property $E$%
. Since $P(s)$ is the minimal property containing $s$, we have $P(s)\subset
E $. Hence $E^{op }\subset P(s)^{op }$ and $t\in P(s)^{op }$. This prove the
inclusion $\{s\}^{\bot }\subset P(s)^{op }$. $\Box $\medskip

In particular, flats of the form $\{s\}^{\bot }$ are properties. Since any
flat is an intersection of subsets of the form $\{s\}^{\bot }$, from the
intersection axiom we obtain that any flat is a property. Thus, we proved
the following important theorem

\begin{theorem}
$\mathcal{P}=\mathcal{F}(\mathbb{S},\bot )$.
\end{theorem}

From Theorem 1 we see that the orthoclosure $A^{\bot \bot }$ of a set $A$ is
the least property containing $A$. It consists of states having the
properties that are common to all elements of $A$. The elements of $A^{\bot
\bot }$ are also called \textit{superpositions} of $A$. The following
Proposition implies that any mixture of $A$ is a superposition of $A$.

\begin{proposition}
Suppose that a state $s$\ is the convex mixture of states $s_{1},...,s_{n}$
with strictly positive coefficients $\alpha _{i}.$ Then the orthoclosure of $%
s$ is the same as the orthoclosure of $\left\{ s_{1},...,s_{n}\right\} $.
\end{proposition}

\textit{Proof.} We have to show that $s$ is endowed with property $P$ if and
only if $s_{1},...,s_{n}$ are endowed with property $P.$ $s\in
P\Leftrightarrow s\left( P\right) =1.$\ But $s\left( P\right)
=\sum_{i}\alpha _{i}s_{i}\left( P\right) .$ Since $s_{i}\left( P\right) \leq
1$ and $\alpha _{i}>0$ for all $i$, we have that $\sum_{i}\alpha
_{i}s_{i}\left( P\right) =1$ if and only if all $s_{i}\left( P\right) =1$
i.e., if and only if $s_{1},...,s_{n}\in P$. $\Box $\medskip

\textbf{Corollary.} \textit{The natural mapping }$\mathcal{F}(\mathbb{P}%
,\bot )\rightarrow \mathcal{F}(\mathbb{S},\bot )$\textit{, where }$\mathbb{P}
$\textit{\ is the set of pure states with the induced orthogonality
relation, is an isomorphism of ortholattices.}\medskip

For this reason we can work with the orthospace $\mathbb{P}$ of pure states
holding in mind that mixtures are in principle possible.

\subsection{Atomicity and the preparation axiom}

A state $s$ is an \emph{atom} if the set $\{s\}$ is orthoclosed. By
Proposition 1, any atom is a pure state. Model 2 from Section 4 shows that
the inverse is not true. Nevertheless in the sequel we restrict our
attention to systems in which any pure state is atom. We formulate this
requirement in terms of measurements.

\begin{axiom}
For any pure state $s\in \mathbb{P}$, there exists a measurement $M$ such
that $\{s\}$ is one of its eigensets.
\end{axiom}

In other words, any pure state is fully characterized by its properties.
Substantively (or operationally) it means that, given any state $s$, there
exists an experimental set-up which can \textquotedblleft
prepare\textquotedblright\ the system in that state $s$. Axiom 3 is rather
reasonable \textbf{(}it is fulfilled in Model 3 from Section 4) and we
explore its consequences. \textbf{\ }

Let $s$ and $t$ be two pure states. Due to Axiom 3, the set $\{t\}$ is a
property and therefore we can speak about $s(t):=s(\{t\})$, the probability
for a transition from the state $s$ to the state-property $t$.

\begin{lemma}
Suppose Axiom 3 is fulfilled. Then $t(s)=0$ if and only if $s\bot t$.
\end{lemma}

\textit{Proof:} Let us suppose that $s\bot t$ and let $P$ be a property such
that $s\in P$ and $t(P)=0$. From the inclusion $\{s\}\subset P$ and the
monotonicity axiom we have $t(\{s\})=0$.\ The converse assertion is more
obvious because $s$ belongs to the property $\{s\}$ on which $t$ vanishes. $%
\Box $\medskip

\begin{corollary}
$s(t)=0$ \emph{if and only if} \ $t(s)=0$.\medskip
\end{corollary}

In the general case, $s(t)$ can differ from $t(s)$.\medskip

Axiom 3 implies the ortho-separability of the orthospace $\mathbb{P}$. Let
us remind that an orthospace $(X,\bot )$ is called \emph{ortho-separable} if
any single-element subset $\{x\}$ of $X$ is a flat. It is easy to check that 
$\{x\}$ is a flat if and only if for any $y\neq x$ there exists $z$
orthogonal to $x$ but not to $y$. For example, the orthospace in figure 3

\begin{eqnarray*}
&\FRAME{itbpF}{2.5953in}{1.663in}{0in}{}{}{qlbfig4.png}{\special{language
"Scientific Word";type "GRAPHIC";display "USEDEF";valid_file "F";width
2.5953in;height 1.663in;depth 0in;original-width 3.3831in;original-height
2.1352in;cropleft "0";croptop "1";cropright "1";cropbottom "0";filename
'/document/graphics/qlbfig4.png';file-properties "XNPEU";}%
}& \\
&\text{Figure\ 3}&
\end{eqnarray*}%
is not ortho-separable, since $a^{\perp \perp }=\{a,c,d\}\neq \{a\}$.\medskip

It is worthwhile noting that any ortho-separable orthospace $(X,\bot )$ can
be reconstructed from the ortho-lattice $\mathcal{F}(X,\bot )$. Recall that
an \emph{atom} of a lattice $\mathcal{F}$ is a minimal non-zero elements of $%
\mathcal{F}$. A lattice $\mathcal{F}$ is called \emph{atomistic} if any
element of the lattice is the join of atoms. If $(X,\perp )$ is an
ortho-separable orthospace then $\mathcal{F}(X,\perp )$ is a complete
atomistic ortho-lattice.

Conversely, let $\mathcal{F}$ be a complete atomistic ortho-lattice. Then,
there exists an ortho-separable orthospace $(X,\perp )$ (unique up to
isomorphism) such that $\mathcal{F}$ is isomorphic to $\mathcal{F}\left(
X,\perp \right) $. One needs to take the set of atoms of $\mathcal{F\ }$as
the set $X$; atoms $x$ and $y$ are orthogonal if $x\leq y^{\perp }$. For
more details see, for example \cite{Mo}. Roughly speaking, ortho-separable
orthospaces and atomistic ortholattices are equivalent objects.

\subsection{Orthomodularity}

It was early recognized that the failure of classical logic to accommodate
quantum phenomena was due to the requirement that the lattice of properties
should satisfy the distributivity law. Birkhoff and von Neumann \cite%
{BirvNeu36} proposed to substitute the distributivity law by the modularity
law. As it turned out, the weaker notion of orthomodularity proved to be
more adequate, see \cite{Kalm}.\medskip

\textbf{Definition.} An orthospace $(X,\bot )$ is said to be \emph{\
orthomodular} if, for every two flats $F$ and $G$ such that $F\subset G$
there exists an element $x\in G$ which is orthogonal to $F$.\medskip

In other words, if $x$ does not belong to a flat $F$ then there exists a
superposition of $F$ and $x$ which is orthogonal to $F$. Orthomodularity
permits constructing orthogonal bases in the same way as in Euclidean
spaces. An \emph{\ orthobasis} of a flat $F$ is a subset $B$ of mutually
orthogonal elements such that $F=B^{\bot \bot }$. In is easy to see that
(for any orthospace) the maximal flat \textbf{1} has an orthobasis. If $X$
is orthomodular then each flat has an orthobasis. More precisely, there holds

\begin{lemma}
Let $F$ be a flat in an orthomodular space, and $B^{\prime }\subset F$ be a
subset consisting of mutually orthogonal elements. Then, there exists an
orthobasis $B$ of $F$ containing $B^{\prime }$.
\end{lemma}

\emph{Proof.} Let $B$ be a maximal (by inclusion) subset in $F$ which
contains $B^{\prime }$ and consists of mutually orthogonal elements. We
claim that the orthoclosure of $B$ coincides with $F$. In opposite case
there exists an element $x$ in $F$ orthogonal to $B$. If we add $x$ to $B$
we extend $B$ which contradicts the assumption of $B$\ being maximal. $\Box$%
\medskip

In particular, beginning with the empty $B^{\prime}$ we can construct an
orthobasis of any flats. Note, however, that ortobases of the same flat
(even the maximal flat \textbf{1}) can have different numbers of elements.

The orthomodularity of space $(X,\bot )$ implies the orthomodularity of the
corresponding ortho-lattice of flats $\mathcal{F }(X,\bot )$. Recall that an
ortho-lattice $\mathcal{F }$ is called \emph{orthomodular} if, for every its
elements $a$ and $b$ such that $a\le b$, the following equality holds 
\begin{equation*}
b=a\vee (b\wedge a^\bot).
\end{equation*}

\begin{lemma}
An orthospace $(X,\bot )$ is orthomodular if and only if its ortholat\-tice $%
F(X,\bot )$ is orthomodular.
\end{lemma}

\emph{Proof.} Let $(X,\bot )$ be orthomodular space and let $F$ and $G$ be
two flats such that $F\subset G$. We have to show that $G=F\vee (G\wedge
F^{\bot })$. Denote by $G^{\prime }$ the right hand side of the expression;
it is clear that $G^{\prime }\subset G$. If the inclusion is strict then $G$
consists an element $x$ orthogonal to $G^{\prime }$. In particular, $x$ is
orthogonal to $F$, that is $x$ belongs to $F^{\bot }$. Since $x$ belongs to $%
G$ as well then $x$ belongs to $G\cap F^{\bot }$ and all the more to $%
G^{\prime }$. But we obtain that $x$ is orthogonal to $x$ which contradicts
the irreflexivity of the orthogonality relation $\bot $. Conversely, let the
lattice $\mathcal{F}(X,\bot )$ be orthomodular and $F\subset G$ be two
different flats. Since $G=F\vee (G\cap F^{\bot })$ then $G\cap F^{\bot }$ is
nonempty. Every element of $G\cap F^{\bot }$ belongs to $G$ and is
orthogonal to $F$. $\Box$\medskip

We now impose the following requirement

\begin{axiom}
Let $P$ and $Q$ be comparable properties (that is either $P\subset Q$ or $%
Q\subset P$). Then there exists a measurement $M\in \mathcal{M}$ such that $%
P=E_{M}(A)$ and $Q=E_{M}(B)$ for some $A, B$ in $O(M)$.
\end{axiom}

This is a serious restriction. For example, it is violated in Model 2 of
Section 3. A first consequence of Axiom 4 is the orthomodularity of the
state space.\medskip

\begin{proposition}
If Axiom 4 is fulfilled then the state space $S$ (or $P$) is
orthomodular.\medskip
\end{proposition}

Indeed, suppose that $P\subset Q$ are two different properties. Let $M$ be a
measurement such that $P=E_{M}(A)$ and $Q=E_{M}(B)$ for $A,B\subset O(M)$.
Obviously, $A\subset B$ and this inclusion is strict. If $b\in B\setminus A$
then every element of $E_{M}(b)$ belongs to $Q$ and is orthogonal to $P$. $%
\Box $\medskip

Another important consequence of Axiom 4 is that any state $s\in \mathbb{S}$
can be considered as a probability measure on the orthospace $\mathbb{P}$%
.\medskip

\textbf{Definition.} A \emph{probability measure} on an orthospace $(X,\bot
) $ is a function $p: \mathcal{F}(X,\bot )\rightarrow \mathbb{R}_{+}$
satisfying the following two requirements:

1) if $F$ and $F^{\prime }$ are orthogonal flats then $p(F\vee F^{\prime
})=p(F)+p(F^{\prime })$;

2) $p(\mathbf{1})=1$.\medskip

By induction we obtain the equality $p(F_1 \vee ...\vee F _n)=p(F_1
)+...+p(F_n )$ for any mutually orthogonal flats $F_1 ,...,F _n$. It is
natural to call this property \emph{ortho-additivity}. The requirement 2) is
simply a normalization. Note that 1) implies $p(\mathbf{0})=0$. If the
orthospace $(X,\bot )$ is orthomodular (what we shall assume) then $p$ is
monotone (that is $p(F)\le p(Q)$ for $F\subset Q$). When all elements of $X$
are orthogonal each to other (and $X$ is a finite set) we come to
conventional notion of a probability measure on $X$, see 5.1.

We already (see 5.1) represented an arbitrary state $s$ as a function on $%
\mathcal{F}(\mathbb{P},\bot )$. We now assert that this function is
ortho-additive.

\begin{proposition}
Axiom 4 implies that any state $s$ (as a function on $\mathcal{F}(\mathbb{P}%
,\bot )$) is a probability measure.
\end{proposition}

\textit{Proof}. Since $s(\mathbf{1})=1$ we have to check the
ortho-additivity of $s$. Let $F$ and $F^{\prime }$ be two orthogonal flats,
and $G=F\vee F^{\prime }$. Since $F\subset G$ then, by Axiom 4, there exists
a measurement $M$ such that $F=E_{M}(A)$ and $G=E_{M}(B)$ for $A\subset
B\subset O(M)$. Set $A^{\prime }=B\setminus A$.

We claim that $F^{\prime }=E_{M}(A^{\prime })$. We begin with inclusion $%
\subset $. Let us consider an arbitrary state $t$ from $F^{\prime }$. The
outcome of the measurement $M$ in the state $t$ cannot belong to $A$
according to orthogonality of $t$ and $F$. Hence the outcome of the
measurement belongs to $A^{\prime }$ with certainty, that is $t\in
E_{M}(A^{\prime })$. Thus, $F^{\prime }\subset E_{M}(A^{\prime })$.

Suppose now that $t$ is a state from $E_{M}(A^{\prime })$, but not from $%
F^{\prime }$. Due to orthomodularity, we can assume that $t$ is orthogonal
to $F^{\prime }$. Since $t$ is orthogonal to $F=E_{M}(A)$ as well, we
conclude that $t$ is orthogonal to $F\vee F^{\prime }=G$ and therefore
cannot belong to $E_{M}(A^{\prime })$. A contradiction. The claim is proven.

Now $s(F\vee F^{\prime})=\mu _M((A\cup A^{\prime})|s)=\mu _M(B|s)=s(G)$. $%
\Box$\medskip

In particular, if $\{b_{1},...,b_{n}\}$ is an orthobasis of a property $F$
then $s(F)=s(b_{1})+...+s(b_{n})$ for any state $s$.

\section{Impact of measurements}

Here we assume that $(\mathcal{M},\mathbb{S},\mu ,\tau )$ is a model of some
measurable system satisfying Axioms 1-4. In the preceding Section we have
shown that the state space $P$ is an ortho-separable orthomodular space. In
this section we show that measurements act as orthogonal projections in this
orthospace.

\subsection{Ideal measurements}

We know that measurements impact on the state, that is the state of a system
is modified by the performance of measurements. Here we investigate
measurements that \textquotedblleft minimally\textquotedblright\ impact on
the state. These measurements are called \textit{ideal}. Let us give a
precise definition. Let $M\in \mathcal{M}$ be a measurement with eigensets $%
F(o)=E_{M}(o),\ o\in O(M)$.\medskip

\textbf{Definition.} A measurement $M$ is \emph{ideal} if, for every state $%
s $, the new state $\tau _{M,o}(s)$ belongs to the convex hull of the flat $%
F(o)\wedge (s\vee F(o)^{\bot })$.\medskip

Note that we earlier said that the transition mapping $\tau _{M,o}$ is
undefined for states belonging to $F(o)^{\bot }$. This is in agreement with
the fact that, for $s\in F(o)^{\bot }$, the flat $F(o)\wedge (s\vee
F(o)^{\bot })=F(o)\wedge F(o)^{\bot }$ is empty. On the contrary, if $s$
does not belong to $F(o)^{\bot }$ then the flat $s\vee F(o)^{\bot }$ is
strictly larger than $F(o)^{\bot }$. According to orthomodularity it
contains an element orthogonal to $F(o)^{\bot }$, that is an element
belonging to $F(o)$. Therefore the flat $F(o)\wedge (s\vee F(o)^{\bot })$ is
nonempty indeed.

\begin{proposition}
Let $M$ be an ideal measurement. Suppose that a pure state $s$ belong to one
of the eigensets of $M$. Then the performance of measurement $M$ leaves the
state $s$ unaffected.
\end{proposition}

In that sense an ideal measurement minimally impacts on states or produces
\textquotedblleft a least perturbation\textquotedblright .\medskip

\emph{Proof.} Let us suppose that $s$ belongs to the eigenset $F=F(o)$. By
the definition of ideality, the new state is in $F\wedge (s\vee F^{\bot })$.
Since $s\in F$ we have the dual inclusion $F^{\perp }\subset \{s\}^{\perp }$%
. By force of orthomodularity $s^{\perp }=F^{\bot }\vee (s^{\perp }\vee F)$.
Applying $\perp $ we obtain the equality $F\wedge (s\vee F^{\bot })=s^{\perp
\perp }$. By Axiom 3, that last set is made out of a single element $s$.
Therefore the new state coincides with $s$. $\Box$\medskip

Let us mention one more property of ideal measurements. When performing an
ideal measurement the new state $s^{\prime }$ is not orthogonal to the old
state $s$. This follows from the fact that $F\wedge (s\vee F^{\bot })$ and $%
s^{\bot }$ do not intersect. In fact, 
\begin{equation*}
s^{\bot }\wedge F\wedge (s\vee F^{\bot })=(s\vee F^{\bot })^{\bot }\wedge
(s\vee F^{\bot })=0.
\end{equation*}

Strengthening Axiom 4, we postulate that there exists sufficiently many
ideal measurements. Let $M\in \mathcal{M}$ be a measurement. As we know,
different flats $E_{M}(o)$ are orthogonal to each other. Moreover, the join
of all $E_{M}(o)$ is equal to $\mathbf{1}$. Indeed, if a state $s$ is
orthogonal to $E_{M}(o)$ then $E_{M}(o)\subset s^{\perp }$ and consequently $%
s(E_{M}(o))\leq s(s^{\perp })=1-s(s)=0$; on the other hand, $%
\sum_{o}s(E_{M}(o))=1$. This leads us to the following definition\medskip

\textbf{Definition.} An \emph{Orthogonal Decomposition of the Unit (ODU)} is
a finite family of flats $(F_{i},\ i\in I)$ such that

a. $F_{i}$ and $F_{j}$ are orthogonal if $i\ne j$;

b. $\vee _{i\in I}F_{i}=\mathbf{1}$.\medskip

Thus, for a measurement $M$, the family of eigensets $(E_{M}(o),\ o\in O(M))$
is an ODU. The next \emph{ideality} axiom asserts that any ODU may be
obtained as the collection of the eigensets of some ideal measurement.

\begin{axiom}
For any ODU $(F_{i},\ i\in I)$, there exists an ideal measurement $M\in 
\mathcal{M}$ with the outcome set $O(M)=I$ and the eigensets $E_{M}(i)=F_{i}$%
.
\end{axiom}

Axiom 5 connects ideal measurements with ODUs. This allows to investigate
the central issue of compatibility of measurements which we do next.

\subsection{Compatible measurements}

In Section 3 we informally discussed the notion of compatible measurements.
In order to consider this issue more formally we need to introduce a notion
of commutativity in the orthomodular space $\mathbb{P}$. Two (or more) flats
commute (or are compatible) if they possess a common orthobasis (see 5.5).
More precisely, a family $(F_{i},\ i\in I)$ of flats \emph{commute} if there
exists an orthobasis $B$ of $\mathbb{P}$ and a family $(A_{i},\ i\in I)$ of
subsets of $B$ such that $A_{i}$ is an orthobasis of the flat $F_{i}$.

For example, flats $F$ and $G$ commute if they are comparable or are
orthogonal. One can show that a family $(F_i, \ i\in I)$ of flats commute if
every two member of the family commute.\medskip

\begin{lemma}
Let flats $F$ and $G$ commute. Then $F\wedge (G\vee F^\bot)=F\wedge G$.
\end{lemma}

Indeed, let $B$ be a common orthobasis of $F$ and $G$. That is $F$ and $G$
are the orthoclosure of some subsets $A$ and $C$ of $B$. Then $C\cup
(B\setminus A)$ is an orthobasis of the flat $G\vee F^{\bot }$ and $A\cap
(C\cup (B\setminus A))=A\cap C$ is an orthobasis of the flat $F\wedge (G\vee
F^{\bot })$. On the other hand, $A\cap C$ is an orthobasis of the flat $%
F\wedge G$. $\Box $

Let $M$ and $M^{\prime}$ be two ideal measurements with eigensets $E_{M}(o)$
, $o\in O(M)$, and $E_{M^{\prime}}(o^{\prime})$, $o^{\prime}\in
O(M^{\prime}) $.\medskip

\textbf{Definition.} The ideal measurements $M$ and $M^{\prime }$ are \emph{%
compatible} ( or \emph{commute}) if every $E_{M}(o)$ commutes with every $%
E_{M^{\prime }}(o^{\prime })$.\medskip

We assert that compatible measurements are compatible in the previously
mentioned informal sense, that is performing one of the measurements does
not affect the results from the other measurement. Indeed, suppose that a
state $s$ is in an eigenset $F:=E_{M}(o)$ and therefore performing $M$ gives
outcome $o$. Suppose further that we perform measurement $M^{\prime }$ and
obtain an outcome $o^{\prime }$. Then the new state $s^{\prime }$ is in the
flat $G\wedge (s\vee G^{\perp })$, where $G=E_{M^{\prime }}(o^{\prime })$.
All the more, the new state $s^{\prime }$ is in the flat $G\wedge (F\vee
G^{\perp })=F\wedge G$ according to according to Lemma 7. Therefore $%
s^{\prime }$ remain in $F$, and if we perform the measurement $M$ again we
obtain the same outcome $o$.

We next show that two ideal measurements are compatible if and only if they
are ``coarsening''\ of a third (finer) measurement. First a
definition\medskip

\textbf{Definition.} A measurement $M^{\prime }$ is \emph{coarser} than a
measurement $M$ (and $M$ is \emph{finer} than $M^{\prime }$) if every
eigenset of $M$ is contained in some eigenset of $M^{\prime }$.\medskip

In other words, outcomes of $M^{\prime }$ can be obtained from outcomes of $%
M $ by means of a mapping $f:O(M)\rightarrow O(M^{\prime })$. In this case
the eigensets $E_{M^{\prime }}(o^{\prime })$ have the form $%
E_{M}(f^{-1}(o^{\prime }))$.

If $M^{\prime }$ and $M^{\prime \prime }$ both are coarsening of $M$ then
they are compatible. For this aim we have to take an orthobasis common to
all eigensets of the measurement $M$.

Conversely, let $M$ and $M^{\prime }$ be two compatible measurements. Then
there exists an orthobasis $B$ common for eigensets of $M$ and $M^{\prime }$%
. If $B$ is a finite set, then we can take as $M^{\prime \prime }$ the
(complete) measurement corresponding to the ODU $B$. In the general case, we
have to consider the family of flats $(E_{M}(o)\wedge E_{M^{\prime
}}(o\prime ))$, where $o$ runs $O(M)$ and $o^{\prime }$ runs $O(M^{\prime })$%
. Because of compatibility, these flats form an ODU. And we have to take as $%
M^{\prime\prime}$ the corresponding measurement which exists by Axiom 5.

Thus, we proved the following

\begin{theorem}
Ideal measurements $M$ and $M^{\prime }$ are compatible if and only if there
exists an ideal measurement refining both $M$ and $M^{\prime }$.
\end{theorem}

\subsection{Canonical decomposition of a model}

We now show that a model of a measurable system satisfying Axioms 1 to 5 can
be written as the direct sums of its irreducible submodels. The argument
below holds for any orthospace but is of largest interest for
ortho-separable orthospaces.

Let $(X,\bot )\ $be an orthospace. We say that two elements of $X$ are \emph{%
connected} if they can be linked be a chain of pairwise non-orthogonal
(tolerant) elements. This relation is an equivalence relation and therefore
divides the set $X$ into classes of connected elements which we denote $%
X(\omega), \ \omega \in \Omega$. Elements from different connected
components are orthogonal to each other; therefore $X(\omega)$ are flats.
These flats are called \emph{central} or \emph{classical}.

If $F$ is a flat in $X$ then, for any $\omega $, the set $F\cap X(\omega )$
is a flat in the orthospace $X(\omega )$ equipped with the induced
orthogonality relation. Conversely, suppose we have a collection of flats $%
F(\omega )$ in $X(\omega )$, $\omega \in \Omega $. Then the union $F=\cup
_{\omega }F_{\omega }$ is a flat in $X$. In other words, the ortholattice $%
\mathcal{F}(X,\bot )$ is the direct (orthogonal) product of ortholattices $%
\mathcal{F}(X(\omega ),\bot )$.

Let us go back to a model of a measurable system with orthospace $(\mathbb{P}%
,\bot )$. As any orthospace, $\mathbb{P}$ decomposes into connected (or
irreducible) components $\mathbb{P}(\omega ),\ \omega \in \Omega $. Since
these components form an ODU then, by Axiom 4, there exists a corresponding
ideal measurement $C$. Since any state is in some component, Proposition 4
implies that the measurement $C$ affects no state. It is natural to call the
measurement $C$ \textit{classical} and to call the corresponding components $%
X(\omega )$ \textit{classical super-states}. The classical measurement $C$
commutes with any ideal measurement. For this reason, we can without loss of
generality, consider only irreducible models.

\subsection{Axiom of Purity}

The property of ideality of measurements significantly narrowed down the
range of the possible impact of a measurement on the state. We know that as
we obtain the result $o$ the system moves from state $s$ to state $%
s^{\prime} $ belonging to the convex hull of the flat $E_{M}(o)\wedge (s\vee
E_{M}(o)^{\bot })$. If this flat is an atom, the new state is uniquely
determined. But if this flat is not an atom (and that is fully possible) the
new state may be a probabilistic mixture of states in $E_{M}(o)\wedge (s\vee
E_{M}(o)^{\bot })$.

To see this, let us consider the example of a fly in a $3\times 2$ box.
There are two measurements: $LR$ and $FCB$. Suppose the state $F$ is
realized as a probability measure $F(L)=F(C)=F(R)=1/3$ (of course, $F(F)=1$
and $F(B)=0$ ). If we, in the state $F$, perform a measurement with
eigensets $\{L\}$ and $\{L\}^{\perp }=\{C,R\}$ and obtain outcome "not $L$",
we may conclude that the image of $F$ is not a pure state but the
equiprobable mixture of states $C$ and $R$. Such a conclusion is in
agreement with the ideality of the measurement $\{C,R\}$ which sends the
state $F$ into $\{C,R\}\wedge (F\vee \{C,R\}^{\bot })=\{C,R\}\wedge (F\vee
L)=\{C,R\}\wedge \mathbf{1}=\{C,R\}$.

We introduce a last axiom guaranteeing that under the impact of a
measurement any pure state jumps into another pure state. Namely, we
consider the following \emph{axiom of purity }

\begin{axiom}
For any pure state $s\in \mathbb{P}$ and any flat $F$ the flat $F\wedge
(s\vee F^{\bot })$ is an atom of the lattice $\mathcal{F}(\mathbb{P},\perp )$%
.
\end{axiom}

We have introduced above a number of non-trivial axioms. We assert that they
are all compatible with each other. Indeed, Example 2 (A fly in a box) gives
a model satisfying to all axioms. Another example is the so-called Hilbert
space model of Quantum Mechanics.\medskip

\textbf{Example 4 }(continued). Let $H$ be a (finite-dimensional) Euclidean
space, as in Example 4. And let $\mathbb{P}$ be the set of all
one-dimensional vector subspaces of $H$. The orthogonality relation is clear
from Example 4. Any flat is given by a vector subspace $V$ and consists of
one-dimensional subspaces in $V$. Measurements are identified with ODUs.
Suppose that $(V_{i},i\in I)$ is a family of pairwise orthogonal vector
subspaces in $H$ and $\sum_{i}V_{i}=H$, and let $v$ be a (non-zero) vector
in $H$ representing some state $s$. Denote by $v_{i}$ the orthogonal
projections of $v$ on subspace $V_{i}$. Then under impact of the
corresponding measurement the state $s$ moves into the state $v_{i}$ with
probability $\cos ^{2}(\varphi )$, where $\varphi $ is the angle between the
vectors $v$ and $v_{i}$ (the probability is 0, if $v_{i}=0$), and give the
outcome $i$. By the construction, this measurement is ideal. It is easy to
check that all other axioms also are fulfilled.\medskip

\textbf{Remark.} In some sense Example 4 is not only a special case. If the
height of $\mathcal{F}$ is more than 3 then the lattice $\mathcal{F}$ can be
realized as a (ortho)lattice of vector subspaces of some Hermitian space
over some $\ast $-field $K$.\footnote{%
The case of the height 1 is trivial: $\mathcal{F}=\{0,1\}$ and $\mathbb{P}$
consists of single state. The case of the height 2 is of more interest. The
(ortho)lattice $\mathcal{F}=\{0,1\}\cup \mathbb{P}$, and the mapping $%
s\mapsto s^{\bot }$ acts on the set $\mathbb{P}$ as an involution without
fixed points. The case of the height 3 is very intricate and unclear.} The
details can be found in \cite{Belcas81} or in \cite{holland95}. If we
additionally require that the orthospace $\mathbb{P}$ is compact and
connected (as a topological subspace of $\Delta (\mathbb{P} ,\bot )$) then
the field $K$ is the real field $\mathbb{R}$, the complex field $\mathbb{C}$
or the skew field $\mathbb{H}$ of quaternions.

\section{Non-classical models in social sciences: A discussion}

In the last section we want to discuss some of the key properties of general
measurable systems in order to help the reader assess their relevance for
social sciences. We wish to emphasize that this section is highly
explorative and should be viewed as a first step that only aims at opening
the discussion.

When applying the theory of measurable system exposed in this paper to
behavioral and social sciences, the general idea is to view an individual as
a measurable system. She is characterized by her type which encapsulates
information about her preferences, attitudes, beliefs, feelings etc. A
decision situation (a situation such that she must choose an alternative out
of set of alternatives) or a questionnaire is a device that measures her
type. Actual behavior, e.g. the choice made in a decision situation, the
actions taken in a game or the answer given to a questionnaire are
measurement outcomes.

In the Introduction we formulated a question as to whether it is possible to
build an interesting theory about a system that changes when being measured.
We answered by the affirmative when imposing a series of properties on
measurements and on their interaction with the system. The state space
representing the system is then endowed with the structure of an atomistic
orthomodular orthospace and the states are realized as probability measures
on the state space.\ We next propose a psychological and behavioral
interpretation of some of those properties.

\ 

\textit{First kindness}

The first key property of a measurable system is that measurements satisfy
first-kindness. Classical measurement theory (including revealed preference
theory) also relies on such an assumption of repeatability. Some reservation
may be in place. We do propose that a choice be viewed as a measurement
outcome that reveals or more precisely actualizes preferences. But in many
settings the repeated character of a choice changes the decision situation.
Clearly, a repeated interaction in a game situation is not equivalent to a
repetition of one and the same decision situation. The prolific theory of
repeated games amply illustrates this. So the repeatedness we have in mind
pertains to elementary situations.

Compared with standard revealed preference theory, the requirement of
repeatability is limited in two respects. The property applies to a smaller
set of choice experiments. As we illustrated in section 4 not all choice
sets can be associated with a first-kind measurement unless we are dealing
with a fully classical agent. Moreover repeatability is only requested in
two \textit{consecutive} identical measurements. If another (incompatible)
measurement is performed on the system in-between, the initial result may
not be repeated. Therefore although the property of first-kindness is
somehow restrictive it is still far less demanding than the standard
classical assumption. Yet, it is not an innocuous assumption and in
particular it precludes stochastic preferences.\bigskip

\textit{Invariance}

Axiom 1 is an axiom of invariance. In the context of choice theory, it is
related to the principle of procedure invariance assumed in classical
rational choice theory. This principle states that a preference relation
should not depend on the procedure of elicitation. Numerous experimental
studies were made on choice versus pricing to exhibit examples of violation
of procedure invariance. It is beyond the scope of this short comment to
systematically compare the classical concept of procedure invariance with
Axiom 1. We confine ourselves to remarking that Axiom 1 applies to systems
in the same state and to remind that it concerns first-kind measurements
only. In particular, we do not take for granted that the pricing or the
matching procedure which are considered as computationally relatively
demanding (compared with a choice procedure) can be performed as a single
measurement rather than as a sequence of (possibly incompatible)
measurements. \bigskip

Axioms 2, 3, 4 and 5 are axioms which all formulate requirements on the
richness of the set of measurements. Another way to look at them that lends
itself to an interpretation in choice theory is that when the set of
primitive measurements is actually limited - in Model 2 there could only be
4 measurements - the axioms imply that choice-measurements may not all be
incompatible with each other. Indeed, it is immediate to see that these
axioms are fulfilled for compatible measurements as these can be combined
into new measurements satisfying the axioms. For this very reason these
axioms do seem very natural to a classically minded person. In a
non-classical world they do not follow naturally, which is demonstrated by
the fact that these axioms are violated in Model 2. Hence, in a choice
theoretic context, these axioms put a limit on how ``non-classical'' agent
(a behavioral system) is allowed to be. Model 3 satisfies all these axioms
and still allows for so-called ``behavioral anomalies''.\bigskip

\textit{States and types}

The notion of state is closely related to Harsanyi's classical notion of
type which is why we use this term when referring to the state of agents.
The Harsanyi type of an agent is a complete description of her preferences,
beliefs and private information such that it allows predicting the agent's
behavior. By observing past behavior, we learn about an agent's type and can
make finer predictions of her future behavior. In Harsanyi's classical world
an information about past behavior is used to predict future behavior
relying on Bayesian updating. The same holds for any compatible
choice-measurements made on a non-classical agent. But generally (when some
measurements are incompatible) learning is not Bayesian. In Model 3 of
section 4, we saw that the performance of the $ab$ measurement on the agent
in state $s$ $\in \left\{ \left[ c>b>a\right] ,\ \left[ c>a>b\right]
\right\} $ erases information about her preferences what concerns the
ordering between $b$ and\ $c$\ so the agent can choose $b$ in$\ \left\{
b,c,d\right\} $.\ 

As in Harsanyi's model, a non-classical pure type is maximal information
about the agent. But in contrast with Harsanyi, a pure type may still be
dispersed (cf Section 3.4) so knowing the pure type does not allow to
predict behavior with certainty. This is reflected in the structure of the
type space. In Harsanyi's type space types are orthogonal to each other. In
the non-classical model not all (pure) type are orthogonal. Instead,
non-orthogonal states are connected with each other in the sense that under
the impact of a measurement the state of the system can transit from one
state to another. This strongly contrasts with Harsanyi's static model where
the act of choosing (i.e. a measurement of the type) has no impact on the
type only on payoffs. The non-classical type space models a changing agent,
we return to this aspect below.\bigskip

\textit{Incompatible measurements}

In the models of rational choice \ of Section 4 measurements corresponds to
sets of alternatives from which the agent makes a choice. Whichever model we
choose, the existence of incompatible choice-measurements implies that the
agent cannot have a preference order on all items \textit{simultaneously}.
Our theory gives a precise meaning to this impossibility. It means: i) if
the ordering over some subset of items is known (possibly only to the agent)
then his preferences over another incompatible subset is random (dispersed
pure states); ii) as the agent makes a choice in a given choice set his type
(preferences) is being modified (measurements affect the state). A
non-classical agent does not have a fixed type (preferences). The
non-classical model is consistent with the hypothesis that an agent's
preferences are shaped in the process of elicitation as proposed by Kahneman
and Tversky (see Introduction).

Thus, the distinctive feature of the non-classical model namely the
existence of incompatible measurements (or alternatively the existence of
dispersed states) delivers a formulation of bounded rationality as the
impossibility to compare and order all items simultaneously. We view this
formulation as particularly interesting because it is also linked to the
idea that an agent's preferences are ``context-dependent'' (see \cite%
{Katver00}, part 6). Both these themes: the issue of comparability in the
universal set of items and intrinsic contextuality of preferences are
central themes in behavioral and experimental economics.

As for today there is no consensus in Physics about the reasons for the
non-classicality of quantum physical phenomena. There exists however a huge
literature on the subject in epistemology. We do not wish to speculate on
reasons for such phenomena in human behavior. Instead, we note that the
(possible) non-classicality of agents invites social scientists to making
interactions the central object of their investigation. Clearly, game theory
is all about interactions but it retains that the type of agents is
exogenous. This cannot be maintained if agents are non-classical systems.
Instead we ought to make agents' type as (partly) endogenous to interaction. 
\footnote{%
Geanakoplos et al. (1989) pioneered an approach in psychological games where
agent's motivation (utility) depends on other's beliefs and therefore are
endogenous to the interaction.} We are currently investigating simpler game
situations along these lines and we trust that it is a promising avenue of
research.\bigskip

\textit{The Stability of the state}

The minimal perturbation principle (ideality) means that a coarse
measurement leaves unperturbed the uncertainty not sorted out by its set of
outcomes. Axiom 5 demands that there exists sufficiently many such ideal
measurements.

In applications to behavioral sciences, an interpretation is that when asked
to choose out of an initial ``state of hesitation'' (a dispersed pure
state), this hesitation is only resolved so as to be able to produce an
answer but no more. The remaining uncertainty is left \textquotedblleft
untouched\textquotedblright . One way to understand this is to see that when
we assume that a choice measurement is ideal it is as if we assumed that the
individual proceeds by taking the "shortest way" to resolve uncertainty. For
instance, suppose as in Model 3 that we ask an individual to make a choice
out of $\{ a,b,c\}$ and that her initial state is a superposition (see
section 5.3) of all six orders. The minimal perturbation principle entails
that the individual will proceed so as to find the most preferred item
without ordering the 3 items. Uncertainty not resolved by an ideal
choice-measurement is left untouched. Therefore we say that ideality implies
a certain stability of the type of a non-classical agent. This can be
contrasted with the classical assumption of complete stability, i.e.
revealing preferences does not affect them at all. Nevertheless ideality may
turn out a rather demanding property, which should be viewed as an
approximation.

Axiom 6 further precises the impact of measurement. It tells us exactly were
a measurement takes the state. We focus on an informational interpretation.
In a social science context it implies that whatever choice the individual
makes that changes her (pure) type, the new behavioral type encapsulates
maximal information about behavior as did the initial type and by ideality
we know the new state.\footnote{\textbf{Information in a pure state is
maximal in the sense that no new information can be obtained from any
measurement without losing some other information, i.e. information that was
true in the initial state but is no longer true in the new state}.} In a
classical context such a pure state corresponds to a state of complete
information. In a non-classical context we know that there exist dispersed
pure states and so a maximal information type (a pure type) does not
uniquely predict behavior in all circumstances.\bigskip

\textit{Caveat}

In applications to behavioral sciences a less attractive feature of our
framework is that a measurement erases information about the previous type.
We should however recall that a person is expected to be composed of a
number of irreducible systems. The loss of memory only applies locally,
within one (irreducible) sub-system. Even within such a system, when
assuming ideality, memory is fully lost only in the case the measurement is
complete (not coarse).\ Yet, our approach implies that to some extent an
individual's previous choices are not relevant to her current type. She may
recall them but she experiences that she has changed. Implicitly, we assume
that at a higher cognitive level, the individual accepts changes in, e.g.
her tastes, which are not motivated by new cognition.

\section{Concluding remarks}

In this paper, we have described the basic structure of non-classical
measurement theory. The objective has been to investigate, from a
theoretical point of view, whether this framework could be suitable for
describing, explaining and predicting human behavior.

As a non-classical measurable system, an agent is characterized by her type
(preferences, attitudes, beliefs etc...) which changes when she makes a
choice actualizing her type. As a consequence behavior exhibits an
irreducible uncertainty. Yet, as we impose some axioms on the interaction
between measurements and the system, behavior is characterized by sufficient
regularity to allow for predictions. We have argued that some of the basic
axioms and properties that underline the theory can be given a meaningful
interpretation consistent with central themes addressed in psychology,
behavioral and social sciences. We also argued that the distinctive feature
of non-classical measurement theory, i.e. the existence of incompatible
measurements, provides an appealing formulation of bounded rationality.

\end{document}